# EXPERIMENTAL AND MODELING STUDY OF THE AUTOIGNITION OF 1-HEXENE / ISO-OCTANE MIXTURES AT LOW TEMPERATURE

G. VANHOVE[1], R. MINETTI[1]

S. TOUCHARD[2], R. FOURNET[2], P.A. GLAUDE[2], F. BATTIN-LECLERC[2*]

[1] Physico-Chimie des processus de Combustion et de l'Atmosphère (PC2A),

UMR n°8522 CNRS,

Université des Sciences et Technologies de Lille, Bâtiment C11

59655 Villeneuve d'Ascq Cedex (France)

[2] Département de Chimie-Physique des Réactions,

UMR n°7630 CNRS, INPL-ENSIC,

1 rue Grandville, BP 451, 54001 NANCY Cedex, France

Full-length article

Shortened running title : AUTOIGNITION OF 1-HEXENE / ISO-OCTANE MIXTURES

[*] E-mail : Frederique.battin-Leclerc@ensic.inpl-nancy.fr ; Tel.: 33 3 83 17 51 25 , Fax : 33 3 83 37 81 20

Autoignition delay times have been measured in a rapid compression machine at Lille at temperatures after compression from 630 to 840 K, pressures from 8 to 14 bar, $\Phi = 1$ and for a iso-octane/1-hexene mixture containing 82 % iso-octane and 18 % 1-hexene. Results have shown that this mixture is strongly more reactive than pure iso-octane, but less reactive than pure 1-hexene. It exhibits a classical low temperature behaviour, with the appearance of cool flame and a negative temperature coefficient region. The composition of the reactive mixture obtained after the cool flame has also been determined.

A detailed kinetic model has been obtained by using the system EXGAS, developed in Nancy for the automatic generation of kinetic mechanisms, and an acceptable agreement with the experimental results has been obtained both for autoignition delay times and for the distribution of products.

A flow rate analysis reveals that the crossed reactions between species coming from both reactants (like H-abstractions or combinations) are negligible in the main flow consumption of the studied hydrocarbons. The ways of formation of the main primary products observed and the most sensitive rate constants have been identified.

# INTRODUCTION

Fuels usually contain six main classes of organic compounds : paraffins, isoparaffins, olefins, naphtenes, aromatics (including polyaromatic compounds) and oxygenates. Commercial gasoline contains approximately 150 different molecules belonging to these 6 classes of organic compounds. Since the end of the last decade, a great number of experimental and theoretical studies has made it possible to simulate the combustion of model molecules taken individually and to predict the formation of characteristic pollutants, as well as autoignition phenomena. Nevertheless, very few studies have been devoted up to now to the case of mixtures containing molecules of several classes of organic compounds. The major part of the studies published on mixture oxidation concerns mixtures of a paraffin and an isoparaffin, an ether or an aromatic compound. The n-heptane/iso-octane mixture is the one which has been the most extensively investigated. It has been experimentally studied in a jet-stirred reactor [1], in a shock tube [2] and in a rapid compression machine [3] and several models for its oxidation have been proposed [1, 3, 4]. In the case of paraffin/ether mixtures, the autoignition of a propane/MTBE mixture has been studied in a shock tube [5], while the oxidation of n-heptane/MTBE and n-heptane/ETBE mixtures has been investigated in a jet-stirred reactor [6]; in both cases, a model has been proposed [5, 6]. The effect of the addition of toluene to the oxidation of alkanes has been investigated in the case of n-butane in a plug flow reactor [7] and in the case of n-heptane in a jet-stirred flow reactor [8]. The global conclusion of these studies is that crossed reactions are negligible and that the synergy effect occurring in a mixture is due to reactions involving the pool of small radicals and especially those influencing the ratio between the concentrations of OH and $HO_2$ radicals [1]. These studies have also demonstrated the inhibiting influence of the presence in a mixture of promoters of resonance stabilized radicals such as MTBE or ETBE, the oxidation of which leads to important amounts of isobutene and then of isobutenyl radicals [6], or toluene [7, 8], which gives benzyl radicals.

Isoparaffins are included up to 30 % and olefins up to 18 % in a commercial gasoline. But apart from a preliminary study of the autoignition of a iso-octane/1-hexene mixture in a shock tube [9], the oxidation of mixtures containing these two types of compounds has not yet been much investigated.

This paper presents an experimental and modeling study of the oxidation and the autoignition of a mixture containing iso-octane and 1-hexene, which are model fuels of isoparaffins and olefins respectively. This study could allow us to know if the presence of linear allylic resonance stabilized radicals have a specific effect on the reactivity of a mixture. Iso-octane is a primary reference fuel for octane rating in spark-ignited internal combustion engines with by definition a Research Octane Number (RON) and a Motor Octane Number (MON) of 100, while 1-hexene is more reactive with a RON of 76.4 and a MON of 63.4. The fact that olefins tends to have lower MON than RON means that they are relatively more prone to autoignition at higher temperature as compared to a primary reference fuel of the same RON.

**EXPERIMENTAL**

The rapid compression machine at Lille has been described extensively in former publications [10, 11] so that only essential features relevant to the present work are presented. Initial gas mixtures were prepared in glass flasks from pure gas by the method of the partial pressures. The combustion chamber of the rapid compression machine is a cylinder of 40 cm$^3$ fitted with a pressure transducer and an optical access to a photomultiplier. The compressed gas undergoes oxidation and autoignition in one or two stages according to its reactivity. Provision has been made for preventing any rebound at the end of compression which could inhibit ignition by an adiabatic cooling. Pressure and light emission are measured every 40 microseconds during and after compression until ignition occurs. The temperature at top dead centre ($T_{TDC}$) is varied by replacing part or whole of nitrogen by argon or carbon dioxide so that the heat capacity of the compressed gas changes. It is calculated using the adiabatic core gas model previously described [12] and can be considered as the maximum temperature reached in the combustion chamber at top dead center.. Pressure at top dead centre is varied by modifying the initial charge. The delay times are taken from top dead centre (TDC). For cool flames, the maximum of the light emission profile is taken as the cool flame occurrence time and is confirmed by the inflexion point in the pressure profile. For final ignition, the very sharp rise in pressure is taken as the final ignition time. For final ignition, the very sharp rise in pressure is taken as the total delay. A sampling attachment was used to collect and quench the reaction gas mixture in a low-pressure vessel at a selected time interval between the cool flame event and final ignition. The collected gas mixture was analysed by gas chromatography. The lighter components of the mixtures were separated with a 50 m porous polymer capillary column and the heavier ones with a 50 m 5 % phenylmethylsilicone capillary column. For the qualitative analyses, a mass spectrometer detector with ionisation by electron impact was used, for the quantitative determinations a flame ionization detector and a thermal conductivity detector. Permanent gases were separated with a 5A molecular sieve capillary column and determined by the thermal conductivity detector. Each species was identified by a thorough study of its spectrum and a comparison with published spectra and/or analysis of the unimolecular fragmentation paths. For quantitative determinations, 4 % neon was added to the initial mixture as an internal standard. The detectors were carefully calibrated for each species or group of species.

**DESCRIPTION OF THE MODEL**

The proposed model has been generated using the EXGAS system and includes some crossed reactions, which are detailed later in the text. The mechanisms generated by EXGAS for the oxidation of 1-hexene [13] and iso-octane [4] taken individually have been previously validated in the same temperature range as the present study.

*General features of the EXGAS system*

As several papers already described the way EXGAS generates detailed kinetic models for the oxidation of alkanes [4, 14, 15] and alkenes [13, 16-18], only a summary of its main features is given here. The system provides reaction mechanisms made of three parts:

♦ A $C_0$-$C_2$ reaction base, including all the reactions involving radicals or molecules containing less than three carbon atoms [19].

♦ A comprehensive primary mechanism including all the reactions of the molecular reactants, the initial organic compounds and oxygen, and of the derived free radicals.

Table 1 summarizes the different types of generic reactions, which are taken into account in the primary mechanism of the oxidation of alkanes and alkenes and shows which reactions are considered for both types of reactants and which are only taken into account in the case of alkenes. Previous papers [4, 13-18] describe in details these generic reactions and the related kinetic parameters.

♦ A lumped secondary mechanism, containing reactions consuming the molecular products of the primary mechanism which do not react in the reaction base.

To ensure the primary mechanism to be comprehensive, all the isomers of reactants, primary free radicals and molecules are considered. Therefore, during the generation of the primary mechanism, both EXGAS and the connected softwares use the detailed chemical formulae of all these molecules and free radicals, reactants and products. To reduce the number of reactants in the secondary reactions, which are the molecular products of the primary mechanism, a lumping of these primary molecules is achieved according to the following rules. Molecules, which have the same molecular formula, the same functional groups and include rings of the same size, are lumped into one unique species, without distinguishing between isomers. This lumping preserves the mass balance and the reactivity of the primary species.

Thermochemical data for molecules or radicals are automatically calculated and stored as 14 polynomial coefficients, according to the CHEMKIN II formalism [20]. These data are computed using the THERGAS software [21], based on group or bond additivity and thermochemical kinetics methods proposed by Benson [22].

Kinetic data are either calculated using the KINGAS software [23] based on thermochemical kinetics methods [22] or estimated through quantitative structure-reactivity relationships [4, 13-18].

*Crossed reactions*

The software EXGAS is designed to consider systematically the crossed reactions: each time a radical is created, it is submitted to all possible generic propagations without considering its reactant of origin. The possible crossed reactions are of two types:

♦ Metatheses involving the abstraction of a hydrogen atom from iso-octane or 1-hexene by a peroxy radical.

In the mechanisms concerning a single class of hydrocarbon, metatheses are only considered for radicals which cannot decompose by beta-scission involving the breaking of a C-C or a C-O bond, apart from the peroxy radicals deriving directly from the reactant (i.e. those which can be obtained through only 2 elementary steps: an H-abstraction or an addition of OH to the double bond followed by an addition to an oxygen molecule). Applying this rule in the present case, only peroxy-iso-octyl, peroxy-1-hexenyl or peroxy-hydroxyhexyl radicals can react by these reactions :

$$1\text{-}C_6H_{12} + C_8H_{17}OO \rightarrow C_8H_{17}OOH + C_6H_{11}$$

$$C_8H_{18} + C_6H_{11}OO \rightarrow C_6H_{11}OOH + C_8H_{17}$$

$$C_8H_{18} + C_6H_{12}OHOO \rightarrow C_6H_{12}OHOOH + C_8H_{17}$$

The abstractions of vinylic H-atoms by peroxy radicals are neglected. The values of the corresponding rate constants are shown in Table 2.

- ♦ The combinations of the allylic radicals deriving from 1-hexene, 1-hexen-3-yl radicals and 3-hexen-2-yl radicals, with $tC_4H_9$ and $iC_3H_7$ radicals.

Amongst the radicals obtained from the decomposition by beta-scission of iso-octyl or deriving radicals, only $tC_4H_9$ and $iC_3H_7$ radicals are taken into account for combinations because they cannot decompose by a beta-scission breaking a C-C bond and are consequently less reactive than other radicals. The rate constants are calculated by KINGAS [23].

The full kinetic mechanism, which includes 5435 reactions involving 1231 species, is available on request. Simulations have been performed using the SENKIN software of CHEMKIN II [20].

**RESULTS AND DISCUSSION**

Figure 1 displays the experimental and simulated delay times obtained for cool flame and autoignition for three different initial pressures in the rapid compression machine. While several experimental studies have demonstrated that the details of the heat transfer in a rapid compression machine are quite complex, both with respect to the geometry and over the time history of the experiments, and that some extents of reaction could occur during compression stroke, [24, 25], satisfactory modeling of ignition delay times measured using the apparatus used in this study has been obtained for a wide range of alkanes from $C_4$ to $C_{10}$ assuming adiabaticity and with conditions after compression calculated based on a core gas model [4, 26]. Figure 1 shows that the autoignition and cool flame delay times are globally well reproduced, apart from the autoignition delay times for the lowest pressure range which are underestimated by a factor about 2 in a wide part of the temperatures range.

Figure 2 presents a comparison between the experimental ignition delay times measured for the iso-octane/1-hexene mixture studied here and those obtained for pure iso-octane and pure 1-hexene, respectively, in the same range of pressure (11.6-14 bar) in the same apparatus. The results concerning iso-octane are those previously measured [27], the values concerning 1-hexene have been extrapolated from previous measurements at lower pressure (7-11 bar) [28] by correlating the logarithm of the ignition delay time with the logarithm of the pressure. The experimental reactivity of the mixture lies globally between that of the two pure compounds, but, at low temperature, it is closer to that of 1-hexene, while it is a little more intermediate to both fuels above 700 K. Considering the differences in MON and RON, it is not surprising that 1-hexene become still more reactive compared to iso-octane at higher temperature. Simulations reproduce well this behaviour, even if the reactivity of iso-octane is slightly overpredicted in the beginning of the negative temperature coefficient area. This figure also shows that simulations with a mixture containing 18 % iso-octane and 82 % 1-hexene lead to the same ignition delay times than with pure 1-hexene in all the temperatures ranges.

Figure 3 displays the experimental and simulated conversions of the reactants vs. residence time for an initial temperature of 707 K in the conditions of figure 1b. This figure also presents the simulated temporal profile of OH radicals mole fraction. In agreement with figure 1b, experiments and simulations lead to close times for the first rise of the conversion of iso-octane and 1-hexene and of the OH radicals mole fraction, showing a good prediction of the cool flame delay time. After the cool flame, while the relative consumption of the two reactants is correctly modelled, the simulated overall reactivity is strongly overestimated inducing to a much shorter ignition delay time than what is experimentally measured.

Despite this disagreement in the global reactivity, it is interesting to compared the experimental and simulated products distributions: Table 3 presents the experimental and simulated selectivities of the different products which have been analysed. These analyses have been performed for a residence time of 19 ms corresponding to the period just after the cool flame in the conditions of Figure 3, for which the experimental conversions of iso-octane and 1-hexene are equal to 24 % and 33 %, while the simulated values are 53 % and 72 %, respectively. The simulated temperature reaches 857 K, but, because of the lower reactivity, the experimental temperature is probably much lower. The selectivity, $S_i(t)$, in % molar of a species i, at a residence time t, is given by the following equation :

$$S_i(t) = 100 \times \gamma_{it} \times n_i(t)/(\gamma_{r0} \times n_r(t_0) - \gamma_{rt} \times n_r(t))$$

with :     $n_i(t)$: Number of mole of the species i at the residence time t,

$n_r(t_0)$: Sum of the numbers of mole of the two reactants at the initial time $t_0$,

$n_r(t)$: Sum of the numbers of mole of the two reactants at the residence time t,

$\gamma_{it}$: Number of atoms of carbon in the species i at the residence time t,

$\gamma_{r0}$: Average number of carbon atoms in the two reactants at the initial time $t_0$, here 7.64,

$\gamma_{rt}$: Average number of carbon atoms in the two reactants at the residence time t.

Hexadiene and one isomer of dimethyl pentenes are not separated in our analysis. The sum of the selectivities of the analysed products is around 50 %, which correspond to a material balance in global carbon around 87 %. That is due to the presence of formaldehyde (the simulated selectivity is equal to 5.3 %) and carbon dioxide (the simulated selectivity is equal to 1.7 %), which have not been quantified, and to the formation of unknown products. No product which can be attributed to a crossed reaction between both reactant has been detected.

Table 3 shows also that the model reproduces the major trends of the distribution of products. Nevertheless, important discrepancies are obtained for the formations of iso-octenes, oxirans, ethylene, carbon monoxide and some oxygenated products obtained in the secondary mechanism (propenol, methylpropenol, butenals…). Quantitative discrepancies may be due to the gradients of temperature which are observed inside the combustion chamber of a rapid compression machine [29] and from the ill-defined temperature history arising from this spatial inhomogeneity. The low formation of $C_4$ ethers is probably due to a too low rate constant of formation, but the value that we use (k = $9.1.10^{10}$ exp(-8380/T) s$^{-1}$) has been shown to be a sensitive parameter for the modeling of ignition delay times for alkanes [4] and is close to the value (k = $7.5 \times 10^{10} \exp(-7675/T)$ s$^{-1}$) proposed by Curran et al. [30] for the same type of reactions. Ethylene and carbon dioxide are usually overestimated by the models obtained from EXGAS, due to the structure of the lumped secondary mechanism. In this part of the model, the molecular products obtained in the primary mechanism are lumped according to their molecular formulae and functional groups and the reactions of these lumped molecules are not elementary steps, but global reactions which result in the formation of molecules or radicals whose reactions are included in the $C_0$-$C_2$ reaction base in the smaller number of elementary steps [19]. This greatly favors the formation of ethylene compared to that of heavier alkenes and that of carbon dioxide compared to heavier oxygenated products. This is an important weakness of the system EXGAS which needs to be improved in the future, but it is difficult to find a better way to accurately reproduce the reactions of the numerous molecular products obtained in the primary mechanism without an exponential increase of the size of the mechanism. However, this approximation does not affect too much the calculations of autoignition delay times.

Figure 4 displays a sensitivity analysis performed by dividing by a factor 10 the rate constant of each generic reaction. This figure shows that all the influential reactions are included in the primary mechanism, except for the decomposition of hydroperoxide species, which has a strong promoting effect at 650 K. Compared to a similar analysis performed in the case of n-heptane [4], this figure shows the influence of initiations which are easier with alkenes, OH additions and combinations. Figure 5 and 6 present flow rate analyses at 857 K in the conditions of Table 3 for the consumption of iso-octane and 1-hexene, respectively. Sensitivity and flow analyses show that the influence of the considered crossed reactions are negligible.

Figure 5 shows that iso-octane reacts by H-abstractions to give the four possible iso-octyl radical isomers. These radicals react with oxygen molecules to give peroxy radicals or to form $HO_2$ radicals and the two isomers of iso-octene shown in Table 3. Iso-octyl radical isomers also decompose to give dimethyl-pentene and methyl radicals, isobutene and iso-butyl radicals, and propene and neo-pentyl radicals. Methyl radicals react with oxygen to form formaldehyde, iso-butyl radicals react with oxygen to give iso-butene and neo-pentyl radicals decompose to give isobutene and methyl radicals. Dimethyl-pentene, propene and isobutene are important primary products as pointed out in Table 3. Iso-octyl peroxy radicals react by isomerization and lead to the formation of ketohydroperoxides or of all the cyclic ethers (from three to six membered cycles) listed as primary products in Table 3.

While bimolecular initiations involving 1-hexene have a slight promoting effect as pointed out in Figure 4, Figure 6 shows that this alkene mainly reacts by addition of OH radical, abstraction of an allylic hydrogen atom and abstraction of a secondary alkylic hydrogen atom. A more detailed flow analysis points out that 4 % of 1-hexene reacts with $HO_2$ to give 2-butyloxirane, 7% reacts by addition of hydrogen atoms to the double bond to give hexyl radicals, 6 % by abstraction of a primary alkylic hydrogen atom and 3 % by abstraction of a vinylic hydrogen atom. The hydroxyhexyl radicals obtained by addition of OH radicals to the double bond and their subsequent isomerizations react mainly by oxidation to produce hexanal and by addition to oxygen to produce $C_6H_{12}OHOO$. These peroxy radicals mainly lead to the formation of formaldehyde, pentanal, and •OH, due to isomerizations/decompositions [31], or to hydroxyhydroperoxyalkyl radicals, which lead ultimately to the production of hydroxyketohydroperoxides or cyclic ethers bearing an alcohol function. The radicals obtained from the abstraction of an allylic hydrogen atom and the following isomerization react mainly by addition to oxygen and lead ultimately to the formation of ketohydroperoxides or unsaturated cyclic ethers, amongst which 2-ethyl-2,5-dihydrofuran, which was analysed during the oxidation of pure 1-hexene in similar conditions [28]. These resonance stabilised radicals can also react by combination with $HO_2$ radicals and form a hydroperoxide, the decomposition of which leads to acroleine. That explains the a priori surprising promoting effect of combinations shown in Figure 4 and the increase of the reactivity of 1-hexene when temperature increases shown in figure 2 and in the differences between its MON and RON. The radicals obtained by abstraction of a secondary alkylic hydrogen atom react by addition to oxygen and lead to unsaturated cyclic ethers, but a fraction of them also reacts with oxygen molecules to give $HO_2$ radicals and several isomers of hexadienes (mainly 1,3-hexadiene) which are listed in Table 3.

Amongst the main products shown by the flow analysis of Figures 5 and 6, all the compounds directly deriving from iso-octane have been found in the analysis shown in Table 3, while only acroleine, pentanal and hexadiene have been detected in the case of 1-hexene. That can be due to the fact that the mixture contains only 18 % of 1-hexene, making the products analysis more difficult than when it is the only reactant. Another reason could be that 1-hexene leads to primary products (e.g. unsaturated cyclic ethers, aldehydes) which are less stable than those of iso-octane (e.g. saturated cyclic ethers, alkenes) and which are then more rapidly consumed by secondary reactions.

The difference of reactivity between 1-hexene and iso-octane is due to two main reasons. First, as it is shown by experiments and simulations, 1-hexene is more rapidly consumed than iso-octane. The additions of OH radicals to the double bond and the possible abstractions of 2 secondary allylic hydrogen atoms, 4 secondary alkylic hydrogen atoms and 3 primary alkylic hydrogen atoms in the alkene lead to a higher global rate constant (k = $1.6 \times 10^{13}$ $cm^3.mol^{-1}.s^{-1}$ at 700 K, for the global reaction with OH radicals) than the possible abstractions of 1 tertiary alkylic hydrogen atom, 2 secondary alkylic hydrogen atoms and 12 primary alkylic hydrogen atoms in the branched alkane (k = $1.0 \times 10^{13}$ $cm^3.mol^{-1}.s^{-1}$ at 700 K, for the H-abstractions by OH radicals). In addition, Figure 4 shows that OH additions and H-abstraction have an important promoting effect. Second, as shown in the case of n-heptane/iso-octane mixtures [1], the isomerizations of peroxyradicals and peroxyhydroperoxide radicals are easier in a linear compound than in a branched compound. In addition in the case of alkenes, the isomerizations leading to resonance stabilized radicals are favoured. The fact that isomerizations are slower in the case of iso-octane favors reactions, which compete with the additions of oxygen. These reactions are oxidations (i.e. the reaction between alkyl radicals and oxygen to give conjugated alkenes and $HO_2\bullet$ radicals) and formations of cyclic ethers (competing with the second addition to oxygen). This is in agreement with the analysis of products displayed in Table 3 and showing an important formation of conjugated alkenes and cyclic ethers deriving from iso-octane and only a small amount of acroleine, pentanal and hexadienes deriving from 1-hexene.

It is interesting to note that in Figure 4, the generic primary reactions with the highest inhibiting effect are oxidations (the reactions with oxygen molecules to give alkenes, dienes, aldehydes or ketones and $HO_2$ radicals) and the formations of cyclic ethers and that those with the highest promoting effect are the additions of oxygen molecules and isomerizations. Oxidations have a strong inhibiting effect at 650 K because of the production of the very stable $HO_2$ radicals. These most abundant $HO_2$ radicals can almost only react by termination, mainly to give $H_2O_2$ molecules, which decompose slowly below 800 K. Consequently, the structure of the molecule of iso-octane, which promotes the formation of alkenes and cyclic ethers instead of that of degenerate branching agents, leads to a much lower reactivity of this compound with oxygen than in the case of 1-hexene, even if this alkene leads to some formations of stabilized allylic radicals. During the reaction of a mixture, there is a common pool of free radicals and the oxidations, formations of cyclic ethers and, to a less extend, some other propagation reactions of the hydroperoxyl radicals (QOOH), e.g. beta-scissions of QOOH producing aldehyde or ketones and OH radicals or alkenes and $HO_2$ radicals, and have an effect on the global reactivity, since they prevent an increase of the global concentration of active radicals. Consequently, when temperature increases, the inhibiting influence of the oxidation (competing with the first addition to oxygen) and formation of cyclic ethers (competing with the second addition to oxygen) deriving from iso-octane increases as the reversibility of the additions to oxygen increases and that makes the mixture less reactive as it is shown in Figure 2. For too low concentrations of iso-octane, the inhibiting influence of these reactions is not strong enough to lower the reactivity induced by the presence of 1-hexene.

In summary, the behaviour of the 1-hexene/iso-octane mixture is not much different from that of n-heptane/iso-octane mixture. The presence of the linear $C_6$ allylic resonance stabilized radicals has no specific effect on the reactivity of the mixture. This behaviour is different from that of mixtures with ethers such as MTBE or ETBE [6] or with toluene [7, 8], for which very unreactive resonance stabilized isobutenyl or benzyl radicals are formed and have a more marked retarding effect.

**CONCLUSION**

This paper presents experimental and modeling results concerning the autoignition of a mixture containing 1-hexene and iso-octane, which appears to have a reactivity close to 1-hexene below 700 K and to be less reactive when temperature increases. The analysis of both experimental products and simulated flow reaction rates shows that crossed reactions are negligible and that 1-hexene is more rapidly consumed and gives less stable products than iso-octane. The decrease of the reactivity of the mixture with temperature seems to be due to the increasing importance of the formation of saturated cyclic ethers and of the oxidations giving iso-octene and the very unreactive $HO_2$ radicals, compared to other propagating reactions leading to branching agents. The presence of the linear allylic resonance stabilized radicals seems to have no specific effect on the reactivity of this mixture with a branched alkanes. This conclusion could be different in the case of linear alkanes, such as n-heptane, which are more reactive or in presence of branched allylic resonance stabilized radicals, such as isobutenyl radicals, which could have a more pronounced inhibiting effect.

While the detailed kinetic model proposed here leads to a globally acceptable agreement with the experimental results, discrepancies for the formation for some important primary products such as iso-octenes and oxirans show that some work is still needed to completely understand the oxidation of alkanes and alkenes. In additions, the secondary mechanism would need to be improved in the future in order to predict the range of products formed more accurately.

**ACKNOWLEDGEMENTS**

This work has been supported by the CNRS, the ADEME, PSA Peugeot Citroën and TOTAL.

TABLE 1 : Generic reactions involved in the primary mechanisms generated by EXGAS for the oxidation of alkanes and alkenes.

| Type of reactions | Reactions common to both alkanes and alkenes | Reactions specific to alkenes |
|---|---|---|
| Molecular reactions | | Ene and retro-ene reactions |
| Initiations | Initiations by breaking of a C-C bond<br>Initiations with oxygen molecules | Initiations between two alkenes molecules involving the tranfer of an allylic H-atom |
| Propagations | Additions of $O_2$ to alkyl radicals<br>Isomerizations involving a saturated transition state<br>Decompositions to saturated cyclic ethers<br>Decompositions by beta-scission<br>Oxidations to give alkenes<br>Metatheses involving the abstraction of an alkylic H-atom | Additions to the double bond<br>Additions of $O_2$ to allyl radicals<br>Isomerizations involving an unsaturated transition state<br>Decompositions to unsaturated or hydroxy cyclic ethers<br>Oxidations to give dienes or oxygenated products<br>Metatheses involving the abstraction of an allylic or vinylic H-atom |
| Terminations | Disproportionations involving peroxy and $HO_2$ radicals | Combinations involving allylic radicals |

TABLE 2 : Kinetic parameters for crossed reaction involving an H-abstraction by a peroxy radical. Rate constants are expressed in the form k = A $T^b$ exp(-E/RT), with the units $cm^3$, mol, s, kcal, by H atoms which can be abstracted.

| Type of abstracted H-atoms | Primary H | | | Secondary H | | | Tertiary H | | |
| --- | --- | --- | --- | --- | --- | --- | --- | --- | --- |
| | lg A | b | E | lg A | b | E | lg A | b | E |
| Alkylic H-atoms | 12.30 | 0 | 20000 | 12.18 | 0 | 17500 | 12.18 | 0 | 15000 |
| Allylic H-atoms | | | | 11.70 | 0 | 14550 | | | |

TABLE 3 : Experimental and computed selectivities, $S_i$, of products (% molar) just after the cool flame (simulated T = 857 K) for a residence time of 19 ms, for initial conditions: T = 707 K, P = 10.9 bar (see figure 3).

| Products | Experiments | Simulations |
|---|---|---|
| **Primary products** | | |
| Propene | 2.2 | 1.8 |
| Isobutene | 9.9 | 8.4 |
| Hexadienes | < 1.8 | 0.7 |
| 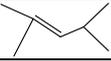 | >1.8 | 2.2 |
| 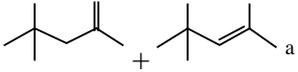 a | 4.1 | 15.2 |
| 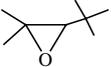 | 0.2 | 3.5 |
| 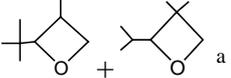 a | 6.7 | 1.1 |
| 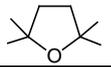 | 10.1 | 14.9 |
| 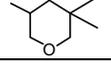 | 1.2 | 0.4 |
| 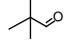 + pentanal | 1.1 | 0.2 |
| **Secondary products** | | |
| Acroleine | 0.8 | 1.4 |
| Propanal | 0.6 | 0.3 |
| Acetone | 2.3 | 5.2 |
| Propenol | 0.04 | 0.01 |
| 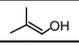 | 0.2 | 0.003 |
| 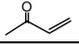 | 0.1 | 0.7 |
| 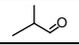 | 1.1 | 1.3 |
| 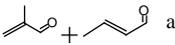 a | 0.8 | 0.1 |
| 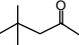 | 0.1 | 0.002 |
| 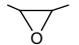 | 0.3 | 0.2 |
| Dimethyl pentadiene | 0.1 | 0.1 |
| 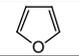 | 0.1 | Not predicted |
| $C_8$ ketones | 1.5 | Not predicted |
| **Products of $C_0$-$C_2$ reaction base** | | |
| CO | 0.3 | 7.0 |
| Methanol | 0.2 | 0.3 |
| Ethylene | 0.8 | 11.0 |
| Acetaldehyde | 1.6 | 2.8 |

a: Products lumped in the simulation results.

**FIGURE CAPTIONS**

Figure 1: Cool flame (white triangles and thin line) and autoignition (black squares and thick line) delay times for iso-octane/1-hexene mixtures containing 82 % iso-octane and 18 % 1-hexene in a rapid compression machine for stoichiometric mixtures and at pressures after compression ranging (a) from 9.4 bar to 10.3 bar, for an initial pressure of 0.59 bar, (b) from 10.3 bar to 12.9 bar, for an initial pressure of 0.72 bar and (c) from 11.4 bar to 13.9 bar, for an initial pressure of 0.79 bar. Symbols are experiments and lines are simulations.

Figure 2: Comparison between autoignition delay times for iso-octane/1-hexene mixtures containing 82 % iso-octane and 18 % 1-hexene (black squares and thick line) and 18 % iso-octane and 82 % 1-hexene (only simulations, dotted line), pure 1-hexene (white triangles and broken line) and pure iso-octane (white circles and thin line) in a rapid compression machine for stoechiometric mixtures and at pressures after compression ranging from 11.4 bar to 14.0 bar. Symbols are experiments and lines are simulations.

Figure 3 : Temporal profiles of the experimental (lines with symbols) and simulated (lines) conversions of iso-octane (black circles and thick line) and 1-hexene (white square and thin line) and of the simulated OH radicals mole fraction (dotted line) at 707 K in the conditions of figure 1b.

Figure 4 : Sensitivity analysis computed under the conditions of figure 1c: the rate constant of each presented generic reaction has been divided by a factor 10. Only reactions for which a change above 10 % has been obtained are presented. To fit in the figure, the changes in % obtained for some very sensitive reactions have been divided by a factor 4 or 12. Underlined generic reactions are those due to alkenes.

Figure 5: Flow rate analysis for the consumption of iso-octane just after the cool flame (at 857 K) for a temperature at the end of the compression of 707 K in the conditions of figure 1c. Compounds in rectangle are shown in Table 3.

Figure 6: Flow rate analysis for the consumption of 1-hexene just after the cool flame (at 857 K) for a temperature at the end of the compression of 707 K in the conditions of figure 1c. Compounds in rectangle are shown in Table 3.

Figure 1

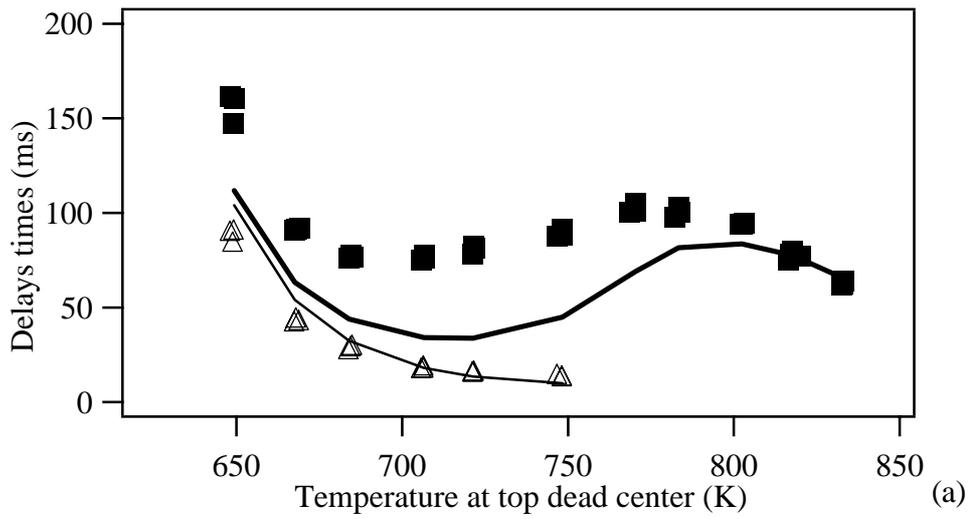

(a)

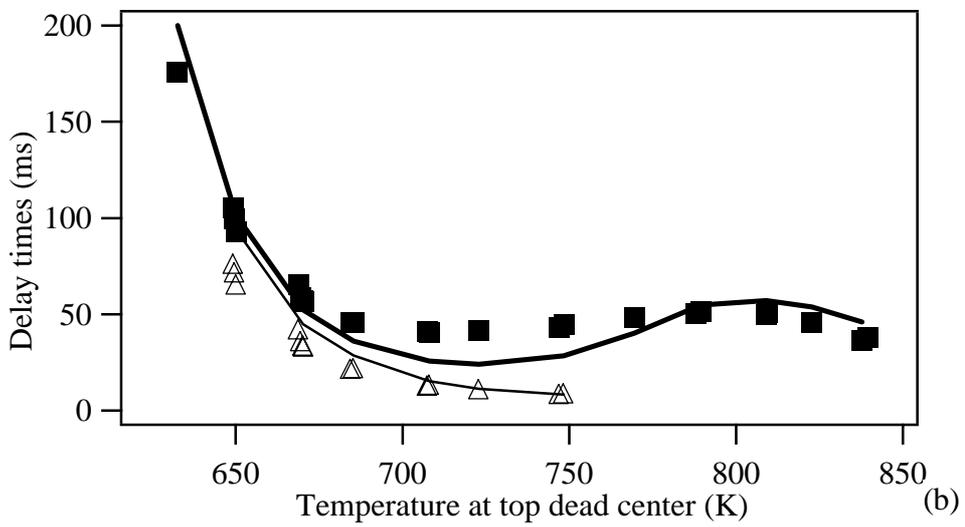

(b)

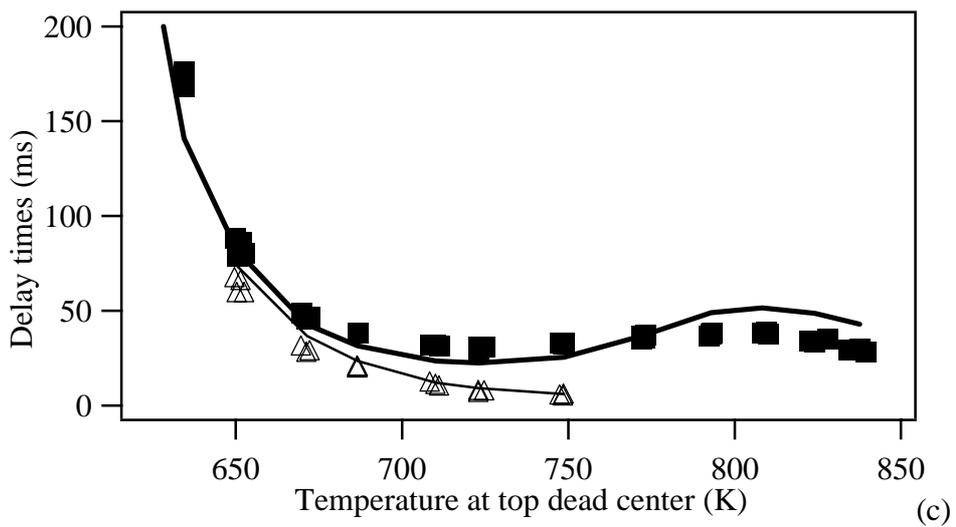

(c)

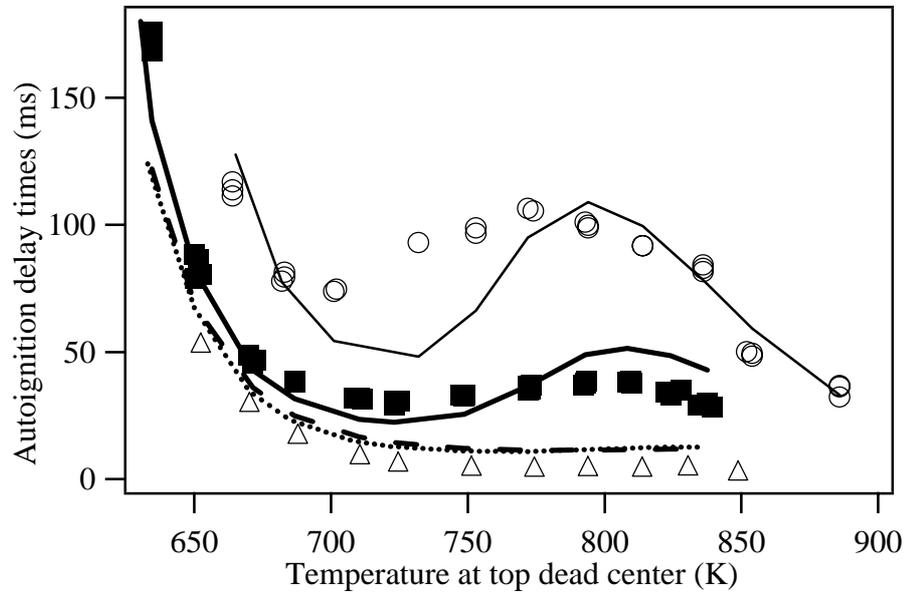

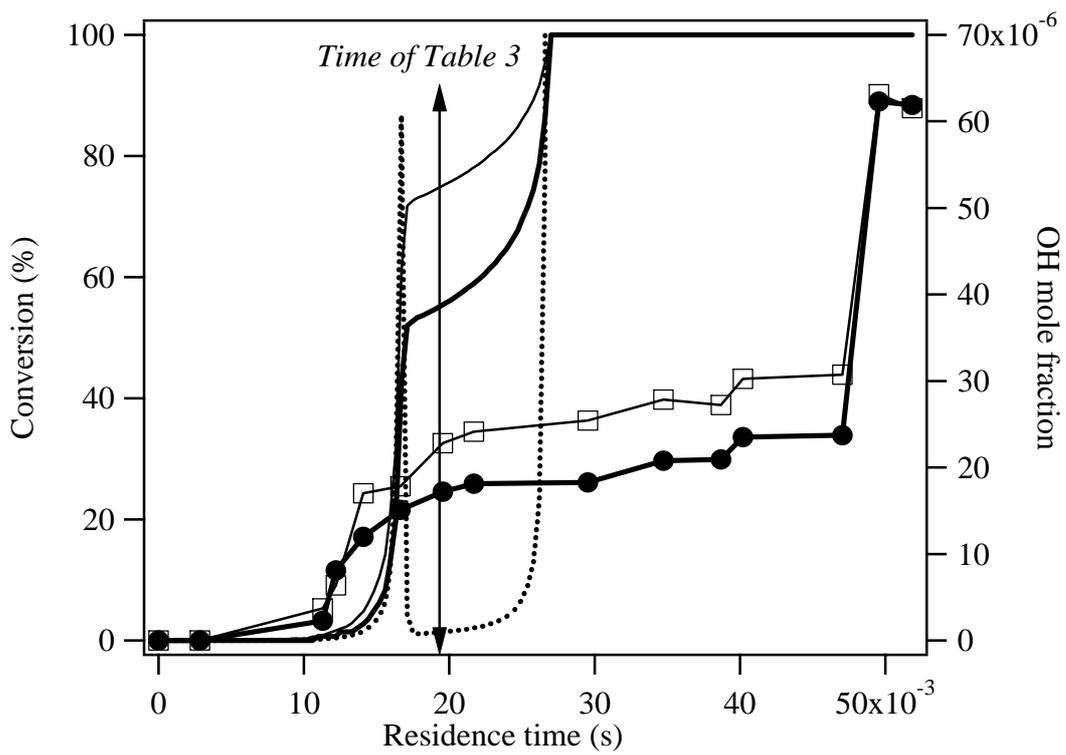

Figure 3

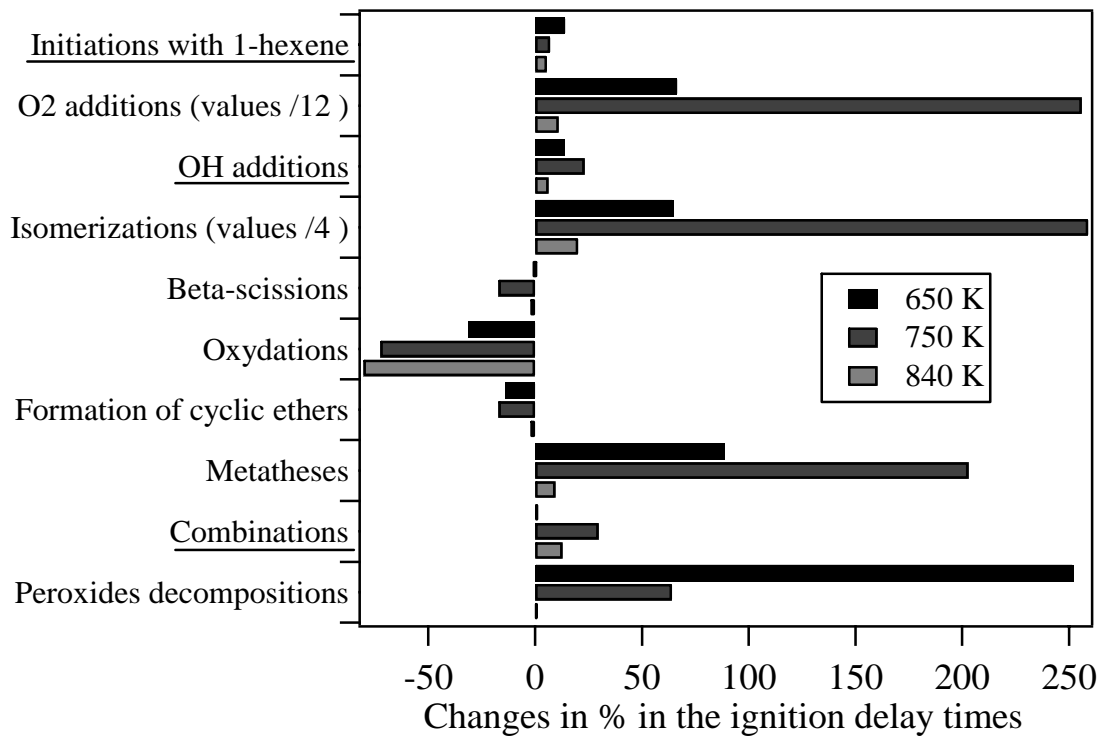

Figure 4

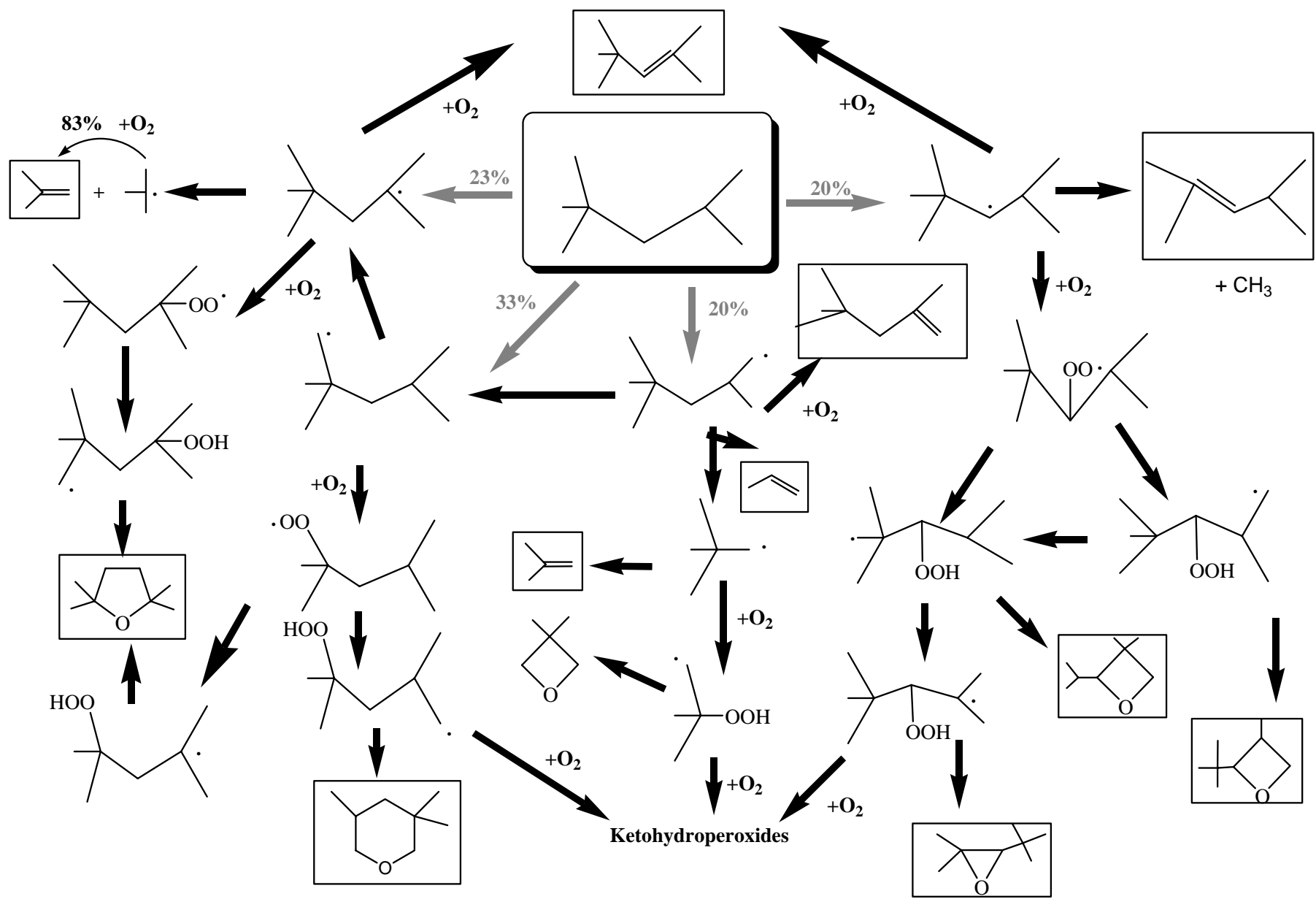

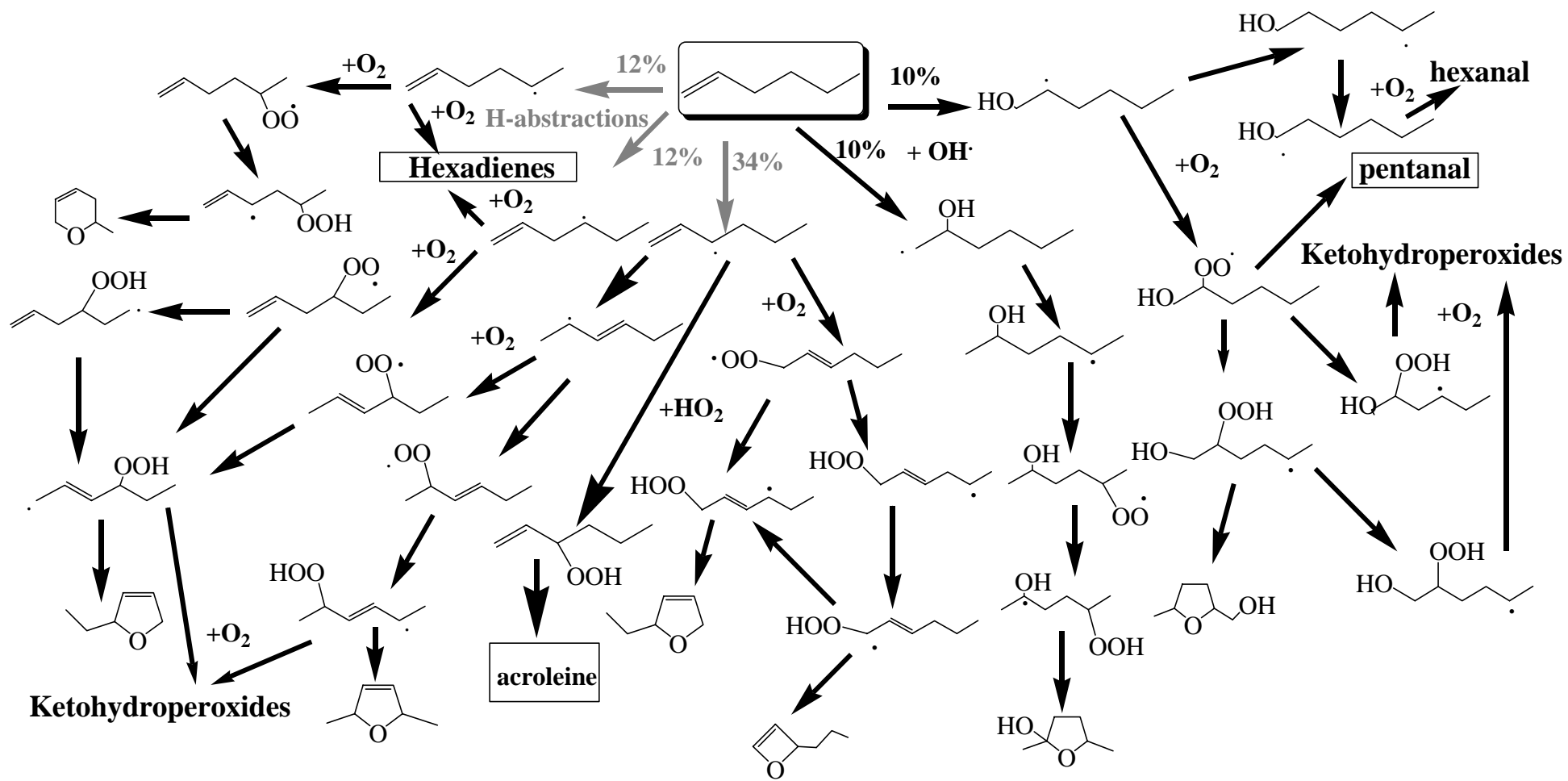